\documentclass[aps,prl,twocolumn,groupedaddress,showpacs]{revtex4}

\usepackage{graphicx}

\begin{document}


\title{Strong Correlations Produce the Curie-Weiss Phase of Na$_{x}$CoO$_2$}


\author{Jan O. Haerter, Michael R. Peterson, B. Sriram Shastry}
\affiliation{Physics Department, University of California,  Santa Cruz, CA  95064}


\date{\today}

\begin{abstract}
Within the $t$-$J$ model we study several experimentally accessible properties of the
2D-triangular lattice system Na$_x$CoO$_2$, using a  numerically exact
canonical  ensemble study of 12 to 18 site triangular toroidal clusters as well as the
icosahedron.  Focusing on the doping regime of $x\sim0.7$, we study
the   temperature dependent specific heat,
magnetic susceptibility and the dynamic Hall coefficient
$R_H(T,\omega)$ as well as the magnetic field dependent thermopower. 
We  find a  crossover  between two phases near $x \sim 0.75$ 
in susceptibility and field suppression of the thermopower arising from
strong correlations. 
An interesting  connection is found  between the
temperature dependence of the diamagnetic susceptibility and the
Hall-coefficient. We predict a large thermopower enhancement, arising
from {\em transport corrections} to the Heikes-Mott formula, in a
model situation where the sign of hopping is reversed from that
applicable to Na$_x$CoO$_2$. 
\end{abstract}


\maketitle

The physics of the 2-dimensional triangular lattice system sodium
cobaltate Na$_x$CoO$_2$ (NCO) is fascinating\cite{nco_general}, combining strong electron
correlations and thermoelectric physics. A Curie-Weiss metallic phase
for dopings $x\sim0.7$ has been reported\cite{ong_nature} where 
the observed physical variables display an unusual mix of behaviors 
that are hybrid between those of good metals and of insulating 
systems challenging theory severely.
The thermopower of NCO is nearly ten times higher than
expected from typical metals; generating excitement in the engineering
and material science communities, particularly applied to
thermoelectric devices. Here we show that strong electron
correlations, along with the geometrically frustrated lattice of NCO,
hold the key to explaining this mysterious state of matter. We examine
several experimentally accessible properties of NCO within the $t$-$J$
model\cite{kumar,baskaran,lee}. We find that strong electron correlations capture the essential
physics and our results compare well with experiment. 

In NCO the low spin  Cobalt ion  valence
fluctuates between a Co$^{4+}$ (spin $1/2$) and a Co$^{3+}$
(spin $0$) configuration; the number of Co$^{3+}$ states is precisely
$x$. The Co ions form a triangular lattice, and 
photoemission\cite{hasan,yang_1} is consistent with a single, hole-like, band with
hopping $t<0$ and $n=1+x$ electrons satisfying the Luttinger volume count.

The $t$-$J$ model describes strongly correlated electron systems by forbidding double 
occupancy of lattice sites. We apply this model to NCO after an electron-hole transformation, requiring $t\rightarrow -t$ and hole doping $|1-n|$. A non-zero $J$ couples nearest neighbor 
electrons via their spin degree of freedom.  For such 
strongly correlated systems perturbation theory is doomed to failure from the 
outset and we make progress through numerical exact diagonalization 
on systems containing 12, 14, and 18 sites on toroidal clusters (periodic 
boundary conditions (BC)) and on ladder clusters (open BC in one 
direction)\cite{long_paper}.  Thermodynamics is considered within 
the canonical ensemble.

The Hilbert spaces of these finite systems 
are very big (up to $\sim$ 80,000 states) and grow exponentially with the number of sites.  
Therefore, all  available symmetries are used  to reduce the dimension of
the matrices that arise to large but manageable proportions.  
However,  Peierls phase factors \cite{sss} are needed to
describe an applied magnetic field,  which breaks or reduces  the translational
invariance, thereby limiting us somewhat. By using a judicious
choice of the BC and of phases on
bonds\cite{kohmoto},  we achieve  a fairly  small non-zero flux per
plaquette of $\pi/N_f$;  where $N_f$ is the total number of
triangular faces on the lattice\cite{long_paper}. The ladder
systems, however, enable an infinitesimal flux to be chosen.  

For NCO, photoemission supports a value for the hopping of $t=-100$ K and we adopt it in this work. This value  is suggested by the ARPES data\cite{hasan}  on the loss 
of coherence of the quasiparticles as well as the dispersion in the composition range $x\sim .7$.  The $T$  dependence of the chemical potential $\mu(T)-\mu(0)$ is another route to estimating $t$ \cite{footnote1}. 

%

Fig. 1(a) shows  the electronic  specific heat  $C_v(T)$, and
is   compared with that  for non-interacting electrons with the same
hopping. We find that the effect of correlations is a shift  in the
peak to a smaller temperature and suppression of  its overall
weight. This is expected since the Gutzwiller projection in the $t$-$J$
model reduces the number of available states and hence the
entropy. Due to a finite system size induced gap in the spectrum, we expect an
exponential behavior of $C_v(T)$ for $T\leq 20$ K, the typical gap
value. Taking this into account, we are  able to extract the linear
electronic contribution $\gamma T$.  The value of $\gamma$ is enhanced
by $\sim 1.5$ over the non-interacting value; this enhancement
depends only slightly on $J$ (neglecting the exponential increase at $T<20K$ due to the finite-system-induced gap) and varies with system size by $0.2$. $J=40$ K (i.e., $0.4|t|$), which is fixed by the experimental system through a comparison of the Curie-Weiss temperature with computations\cite{ong_nature, long_paper}.

\setlength{\unitlength}{1cm}
\begin{figure}
\begin{picture}(0,0)(0,0)
\put(-.6,8.2){{\large (a)}}
\put(-.6,6.5){{\large (b)}}
\put(-.6,5.3){{\large (c)}}
\put(-.6,4.1){{\large (d)}}
\put(-.6,2.8){{\large (e)}}
\put(-.6,1.5){{\large (f)}}
\end{picture} 
\includegraphics[width=7cm]{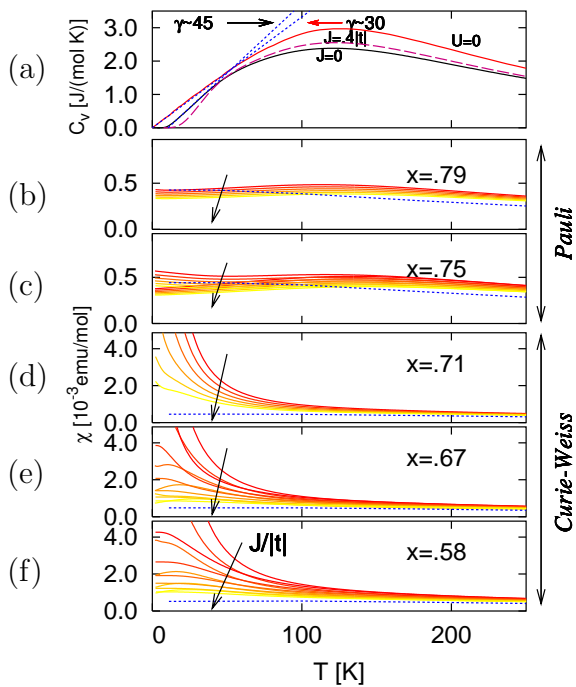}
\caption{(color online) Specific heat and susceptibility. 
(a), specific heat $C_v(T)$ for $x=0.72$, computed on the 18-site cluster,
comparison of $J=0$ (bottom) with $J/|t|=0.4$ (middle) and bare (Hubbard $U=0$)
specific heat (top), dotted straight lines show linear fits and
$\gamma$ values for $J=0$ and $U=0$ in units of $mJ/(mol\cdot$K$^2$). (b)-(f),
susceptibility $\chi(T)$ for dopings around $x \sim 0.7$. The dotted curves
indicate the  bare susceptibility, and  arrows indicate the evolution of
$J/|t|$ from $0$ to $0.5$ in steps of $0.1$ (red to yellow).   Note the  change of scale
in different panels. These results combine two  different clusters, a
12-site torus ($x=0.58$, $0.67$, $0.75$) and a 14-site torus
($x=0.71$, $0.79$).  The difference in  $x=0.71$ and $x=0.75$
shows that  $\chi(T)$ transitions from a Curie-Weiss to Pauli
paramagnetic behavior in this range.}
\end{figure}

In Fig. 1(b)-(f) 
we present the spin susceptibility  $\chi(T)$ for several dopings 
around $x\sim0.7$. In the
band limit $x\rightarrow 1$, as in the upper two panels (as well as 
results not shown), we find the
expected weakly T dependent but $J$ insensitive Pauli paramagnetic
behavior.  When $x$ is lowered below $x=0.75$ (bottom three panels),
$\chi(T)$ shows strong Curie-Weiss-like $T$ and $J$ dependence, and is
significantly renormalised from the non-interacting value at low
$T$. This indicates a crossover to the strong-correlation induced
local moment behavior for  $x < 0.75$  which closely parallels 
experimental findings\cite{ong_nature}. In this Curie-Weiss
phase,     the behavior at high $T$    is  described by the 
Curie-Weiss form
$\chi(T)=\frac{1}{3}\frac{1}{v}\frac{\mu_B^2p_{eff}^2}{k_B(T-\theta)}$
with a negative Weiss temperature $\theta$ and effective magnetic
moment $p_{eff}$, $v=V/N$ is  the unit cell volume. When continuing
the analysis to $x\rightarrow 0$, antiferromagnetic (AFM) correlations increase and we find
that $\theta(x,J)=-c J_{eff}(x)$ where $J_{eff}(x)= J(1+c'' x
|t|) + c' x |t|$, with $c= 4.0$, $c'=0.01425$, and $c''=-0.9175$. The $c'$
term originates in the kinetic antiferromagnetism of the frustrated
lattice\cite{counter_nagaoka}, and signifies that even in the absence
of $J$, there is a tendency for  AFM order, i.e., in a direction
opposite to the usual Nagaoka mechanism for the square
lattice\cite{anderson_nagaoka}. 

Experimentally, the Hall
coefficient  of NCO is  remarkable in many respects.  Most striking is
the unbounded linear increase with temperature of the Hall coefficient
$R_H$. To understand this  we   perform the brute force exact summations
of Kubo's formulae for various conductivities\cite{sss} by
introducing a level width, i.e., a broadening $\omega \rightarrow
\omega + i \eta$ with $\eta$ equal to the mean energy level
spacing. In addition, we   evaluate the high frequency limit\cite{kumar} of 
$R_H$ (called $R^*_H$) for all $T$. Recall that the high $T$
estimates of $R^*_H$  led to a prediction\cite{kumar,sss} of the
linear T dependence of the Hall constant for NCO, which was
successfully verified\cite{wang_hall}. We  are thus able to provide a
purely theoretical benchmarking of this idea as well, subject of
course to  the limitations of the finite size  clusters.

Focusing  on the region of doping around $x\sim 0.7$ Fig. 2(a) 
shows the Hall coefficient as a function of
temperature and frequency. We  find that the Hall coefficient is
relatively insensitive to frequency, in keeping with the original
expectations\cite{shastry_1,kumar,sss}. All curves show a minimum
near $T=100$ K, and an unbounded linear increase for $T>200$ K. The slope of
$R_H^*$ as found in the clusters is in agreement with results from
high-temperature expansions\cite{kumar}. The experimental curve,
unlike theory, has a change of sign and also a pronounced minimum at
$T\sim 100$ K. We cannot reproduce the change of sign. While we could
fit the high $T$ slope  more accurately, it requires a smaller value
of hopping $t \sim 35$ K, as already noted\cite{kumar}. Such a
choice  would make most other variable fits less sensible. Hence we
conclude that while the data\cite{wang_hall} is largely as
expected from theory, in detail it is still not possible to reconcile
the change in sign as well as the magnitude of the  slope with theory.  
Therefore, the Hall coefficient still offers a considerable challenge 
to both theorists and experimentalists.

In an alternate effort  to understand further the data\cite{wang_hall} 
at lower $T\sim 100$ K, we note that the curvature
of the (Landau) diamagnetic susceptibility $\chi_{d}$ (obtained by
inserting Peierls phases)    and the Hall constant are curiously
related in our computation via
\begin{equation}\label{chiL_RH_eq}
T\partial^2\chi_d/\partial T^2=f(x)\partial^2 R_H/\partial T^2\;,
\end{equation}  
where $f(x)$ is a
function depending on doping $x$. We  integrate equation (\ref{chiL_RH_eq}) 
to arrive at the Hall 
constant from our $\chi_{d}$, yielding a  good overall fit of
data. The problem with the slope and the negative
intercept at low $T$ are forgiven in this approach since one fits the
two constants, but it does capture the ubiquitous minimum at $T \sim
100$ K, with a slight    $J$ dependence\cite{long_paper}. While the
Landau diamagnetism is very small compared to the paramagnetic
susceptibility for narrow band systems, it might yet be accessible to
experiments. This is so since  it is anisotropic in space (being
planar), in contrast to the isotropic Pauli susceptibility, and hence
torque magnetometry\cite{torque}  could help disentangle these
terms.

\begin{figure}
\begin{picture}(0,0)(0,0)
\put(-.7,7.){{\large (a)}}
\put(-.7,3.6){{\large (b)}}
\put(-.7,1.5){{\large (c)}}
\end{picture} 
\includegraphics[width=6.7cm]{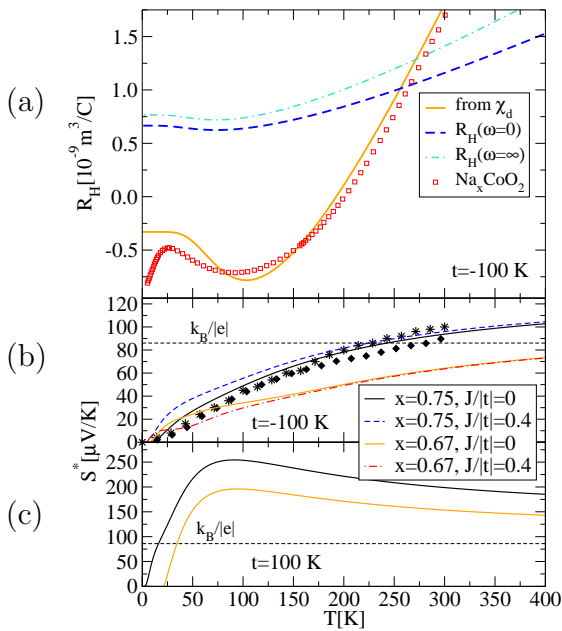} 
\caption{(color online) Hall coefficient and thermopower.  
(a), Comparison of several results for the Hall coefficient $R_H(T)$ at $x=0.75$ with experiment in ref. \cite{wang_hall} at $x=0.71$
(red squares): $R_H^*$ (blue dot-dash), $R_H({\omega}=0)$ (blue dashed), $R_H$
(orange solid) derived from $\chi_d$ (at $x=0.83$, ladder), the dc-limit
required a broadening of the frequency $\omega\rightarrow
\omega+i\eta$ with $\eta\approx 3|t|$ to eliminate finite-size
artifacts.  All results are for $12$-site clusters and $J/|t|=0$.
(b), Infinite frequency thermopower $S^*$ vs. $T$ for a $12$-site torus at $x=0.75$ and $x=0.67$.  
The solid black and
dashed blue lines correspond to $J/|t|=0$, and $0.4$ at $x=0.75$, respectively, while
the solid orange and dashed-dotted red lines correspond to $J/|t|=0$, and $0.4$ at $x=0.67$, 
respectively. $S^*(T)$ for $t=-100$ K relevant for NCO.  The diamonds
and stars represent measured thermopower for NCO at $x=0.68$ from ref.~\cite{ong_nature}
and ref.~\cite{terasaki}. (c), Our prediction for $S^*(T)$ for the case 
when the sign of the hopping is reversed ($t=100$ K.)} 
\end{figure} 

The thermopower of NCO is striking in its large 
magnitude\cite{terasaki} $\sim 100\mu V$/K and also in its surprising sensitivity to an
applied  magnetic field\cite{ong_nature}. A formulation for the
thermopower has been recently given\cite{shastry_1} in the high
frequency limit in the same spirit as the Hall constant.  We note
that the thermopower can be written as a sum of a transport and the
Heikes-Mott term as $S(\omega,T)= S_{Tr}(\omega,T)+ S_{HM}(T)$, where the 
Heikes-Mott term is $S_{HM}(T)= \frac{ \mu(0)- \mu(T)}{q_e T}  $, and 
the transport term is
\begin{equation}
S_{Tr}^*(T)=\lim_{\omega \rightarrow \infty}  S_{Tr}(\omega,T)  = \frac{q_e
(\Delta(T)-\Delta(0))}{T \langle \tau^{xx} \rangle}. 
\end{equation}
Here $q_e = -|e|$ is the electronic charge and $\hat{\tau}_{xx}$  is
the diagonal part of the stress tensor\cite{sss,shastry_1}.  The term
$S_{HM}$ is entropic  in origin\cite{beni_chaikin}. Detailed
expressions for $\Delta$ as an expectation of a many-body operator in
the $t$-$J$ model are  given in equation (83) of ref.~\cite{shastry_1}, and in a
longer work\cite{long_paper}. 

The main approximation in ref.~\cite{shastry_1} is to use the high
$\omega$ limit of $S(\omega,T)$ (called $S^*$), and is expected to be numerically quite
reasonable, in parallel to the behavior of the Hall constant reported
in this work. At low $T$ the two contributions to $S$ vanish
separately as we have written them, and in general we find that
for $t<0$ (the case of NCO) the entropic part is by far the dominant term. For the
opposite case ($t>0$) the transport term comes into play in a dominant
way, and leads to very interesting behavior as we shall show.

We first discuss the  thermopower in the absence of magnetic field.
Fig. 2(b) shows the $T$ and $J$
dependence of $S^*$ for $x$ relevant to the Curie-Weiss phase. The
Heikes-Mott term dominates over the transport term and the 
frequency dependence of the thermopower (evaluated via Kubo formulae) 
is found to be quite weak in this range of doping for  $t<0$ relevant to NCO\cite{long_paper}, and thus the high
frequency approximation is as good as exact.  We find our results 
compare well with the experimental data\cite{terasaki,ong_nature}.
We now sharpen  the prediction
(from Eq. (88) of ref.~\cite{shastry_1}) that reversing the sign of
hopping leads to a maximum in $S^*$ as a function of $T$. Fig. 2(c) 
indeed shows such a behavior and provides an estimate of the
expected enhancement in $S^*$, a factor of 2 to 3 over the frustrated
case of NCO and also of the expected temperature scale $T_{max} \sim
|t|$. It would be interesting to check this experimentally.  A part of the enhancement for $t>0$ arises 
from the prominent peak in the single particle density of states coming close to the 
fermi level. Our numerics show that interactions further amplify this effect 
considerably (a factor of $\sim 2$).

\begin{figure}
\begin{picture}(0,0)(0,0)
\put(-.4,4.){{\large (a)}}
\put(7.6,6.){{\large (b)}}
\put(7.6,3.7){{\large (c)}}
\put(7.6,1.4){{\large (d)}}
\end{picture} 
\includegraphics[width=4.5cm]{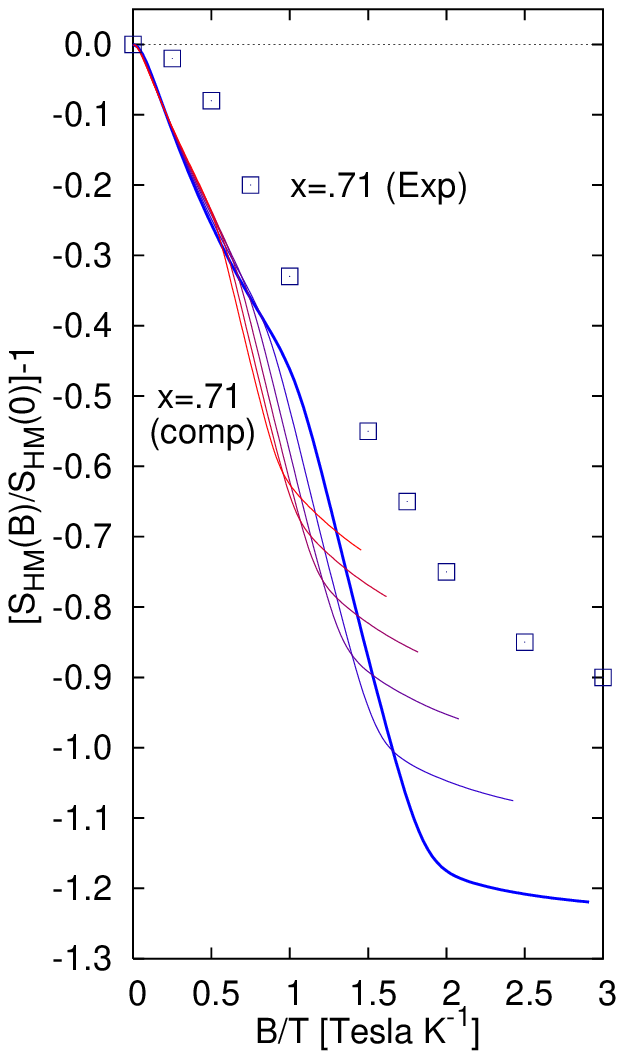}
\includegraphics[width=3cm]{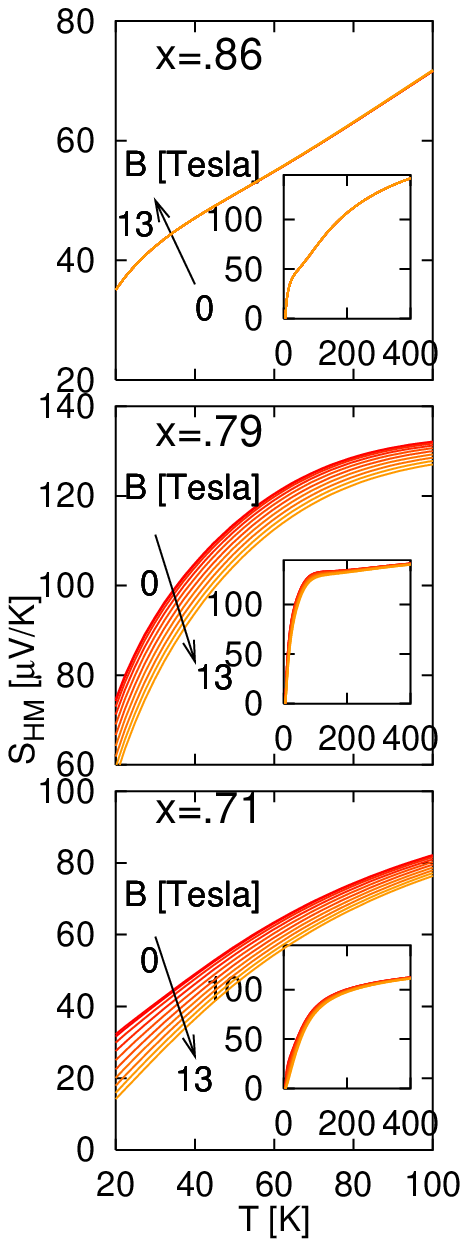} 
\caption{(color online) Magnetic field dependence of the thermopower. (a), $S_{HM}(B)$ normalised to
$S_{HM}(0)$ shows the relative suppression of $S_{HM}$ with $B$
for temperatures $T=(100,120,140,160,180,200)$ K (blue to red), horizontal axis
is rescaled by $1/T$ as done in ref.~\cite{ong_nature}. (b)-(d), $S_{HM}(T)$ for several doping values around $x=0.71$ and several values of 
$B$ from 0 T to 13 T (red to orange).  Note that for
$x>0.75$ the field-dependence becomes very weak.  The insets display the full 
temperature behavior of $S_{HM}$ out to $T=400$ K.  All data for $J=0$.  
Finite $J$ reduces the field suppression.} 
\label{Heikes_T}
\end{figure} 

Next we consider the effects of a  magnetic field 
of strength $B$ (perpendicular to
the plane) on  the thermopower. The field induced change of the
transport term arises from the Peierls factors, and is found to 
be a very small fraction ($\sim 0.001$) of the change in the Heikes-Mott term. 
The field dependence of the chemical potential is significant and responsible
for the overall change. In Fig. 3(a) 
we show the normalised $S_{HM}(B,T)$ for several dopings as function of $B/T$. This scaling
with $B/T$ is very similar to the one found in ref.~\cite{ong_nature}, both
qualitatively and quantitatively.   In Fig. 3(b)-(d) we find a crossover for  $x\sim
0.75$ from a weakly $B$-dependent $S(B,T)$  at $x\geq 0.75$  to  the
Curie-Weiss  phase where $S$ is greatly suppressed by $B$. This
crossover is similar to the one seen in the $T$ dependence of the
spin-susceptibility.  These results confirm   the interpretation\cite{ong_nature} in terms
of spin-entropy as the leading contribution to the field-suppression
and also provide a guide to what one can expect at high magnetic
fields that are not accessible experimentally.

\begin{figure}
\begin{picture}(0,0)(0,0)
\put(-0.75,-1.3){{\large (a)}}
\put(-0.75,-3.5){{\large (b)}}
\end{picture}
\includegraphics[width=5.0cm,angle=-90]{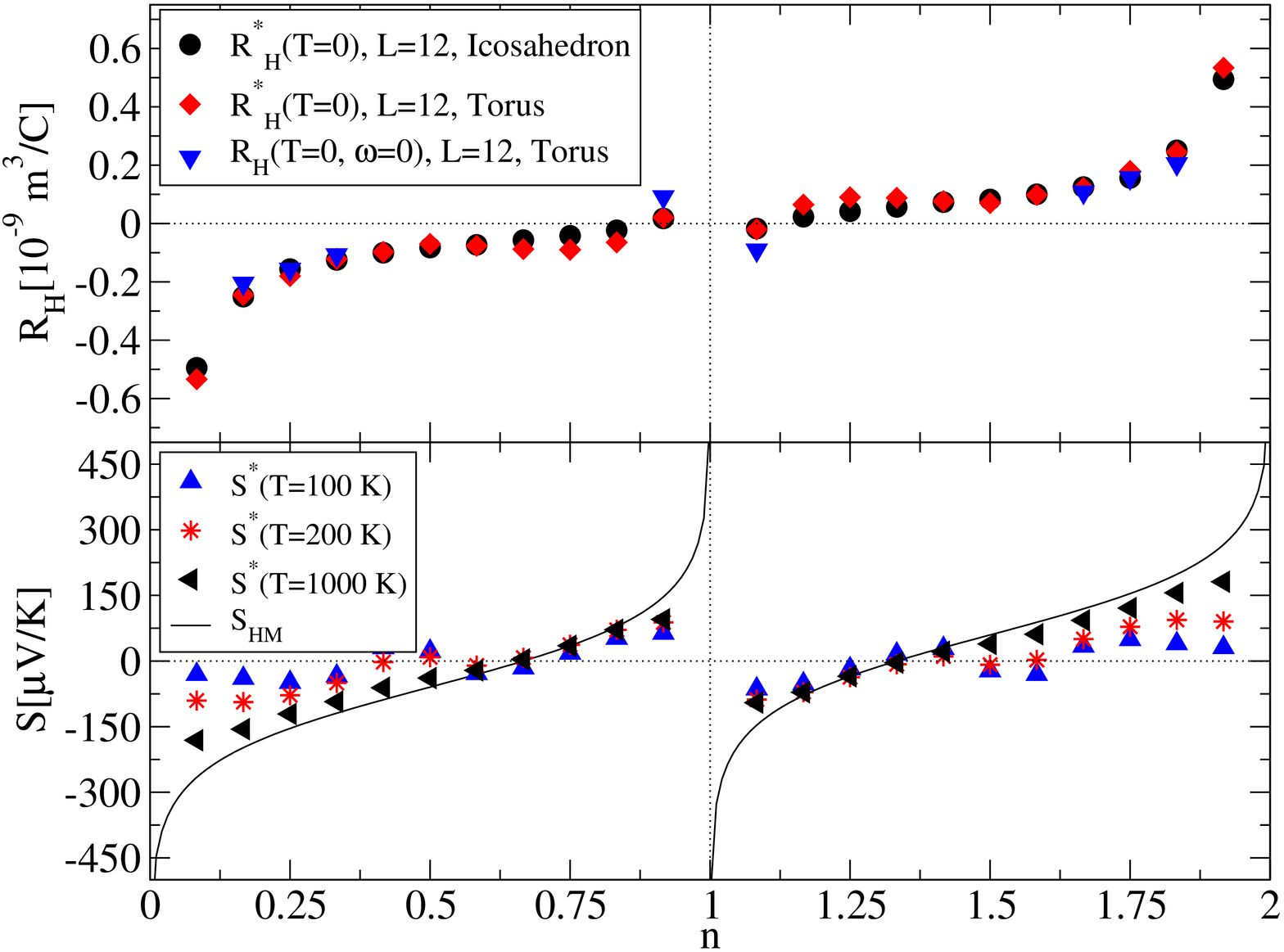} 
\caption{(color online) Hall coefficient and thermopower as a function of filling $n$.  
(a), Filling dependence of $R^*_H(T=0)$ for a $12$-site torus (black 
circles) and icosahedron (red diamonds) as well as
$R_H(T=0,\omega=0)$ for the $12$-site torus (blue triangles) all for
$J/|t|=0$.  (b), Filling dependence of $S^*(T)$
for $T=100$ K, $200$ K, and $1000$ K (blue triangle, red star, black triangle)
for $J/|t|=0$ on a $12$-site torus.  The solid black line is $S_{HM}$ at infinite $T$.  The
dotted lines in both panels are guides to the eye indicating zero
$R_H$ and $S$ and half filling ($n=1$).}  
\label{RH_S_x}
\end{figure} 

We next make a few comments about the overall behavior of the
thermopower and Hall constant as a function of filling in  strong
coupling theories. As noted  earlier\cite{sss,phillips}, while band
theory predicts a single zero crossing of the Hall constant as  $n$
increases from $0 \rightarrow 2$, a strongly correlated electron
system has three crossings.  In precisely the same sense the
thermopower also has three zero crossings for a correlated band
system. We demonstrate this for the triangular lattice.  In
Fig. 4(a)-(b) we show the filling dependence of both the Hall
constant and the thermopower. The divergence of the Hall constant at
half filling is forced by the carrier freeze out accompanying the Mott-Hubbard gap, 
and is less pronounced in the triangular lattice than in
the square lattice\cite{sss}. For the thermopower we expect similar
results for the square lattice.

The hopping parameter $t$ ($\sim 100$ K) used  here to reconcile 
a considerable body of experimental data with theory, is much smaller
than that in   high $T_c$ systems $t \sim 3000$ K.  This value  
is also smaller, by a factor of 10, than the one found in LDA studies \cite{singh,pickett},
and should motivate further studies of the difficult problem of
projecting a multi-band system onto a single band model
systematically. In conclusion, our work shows
that  strong correlations  account for  the observed dramatic behavior
of the Curie-Weiss metallic phase in considerable quantitative detail.

\begin{acknowledgments}
We thank N. P. Ong and Y. Wang for valuable discussions. This work is 
supported by Grant No. NSF-DMR0408247.
\end{acknowledgments}


\end{document}